# Room temperature strong orbital moments in perpendicularly magnetized magnetic insulator

*Ganesh Ji Omar, Pierluigi Gargiani, Manuel Valvidares, Zhi Shiuh Lim, Saurav Prakash, T. S. Suraj, Abhijit Ghosh, Sze Ter Lim, James Lourembam[*], Ariando Ariando[*]*

Dr. G. J. Omar, Dr. S. Prakash, Dr. T. S. Suraj, Prof A. Ariando
Department of Physics, National University of Singapore, Singapore 117542, Singapore
Email: ariando@nus.edu.sg

Dr. P. Gargiani, Dr. M. Valvidares
CELLS-ALBA Synchrotron Radiation Facility, Cerdanyola del Valles, Spain

Dr. Z. S. Lim, Dr. A. Ghosh, Dr. S. T. Lim, Dr. J. Lourembam
Institute of Materials Research and Engineering, A*STAR (Agency for Science, Technology and Research), 2 Fusionopolis Way, Innovis, Singapore 138364, Singapore
Email: james_lourembam@imre.a-star.edu.sg

**The balance between the orbital and spin magnetic moments in a magnetic system is the heart of many intriguing phenomena. Here, we show experimental evidence of a large orbital moment, which competes with its spin counterpart in a ferrimagnetic insulator thulium iron garnet, $Tm_3Fe_5O_{12}$. Leveraging element-specific X-ray magnetic circular dichroism (XMCD), we establish that the dominant contribution to the orbital moment originates from 4$f$ orbitals of Tm. Besides the large Tm orbital moment, intriguingly, our results also reveal a smaller but evident non-zero XMCD signal in the O $K$ edge, suggesting additional spin-orbit coupling and exchange interactions with the nearest neighbour Fe atoms. The unquenched orbital moment is primarily responsible for a significant reduction in $g$-factor, typically 2 in transition metals, as determined independently using ferromagnetic resonance spectroscopy. Our findings reveal a non-linear reduction in the $g$-factor from 1.7 at 300 K to 1.56 at 200 K in $Tm_3Fe_5O_{12}$ thin films. These results provide critical insights into the role of the $f$ orbitals in long-range magnetic order and stimulate further exploration in orbitronics.**

## 1. Introduction

Orbital angular momentum, $\boldsymbol{L}$ is a fundamental degree of freedom for understanding many-body physics in solids. $\boldsymbol{L}$ determines the strength of the orbital moment, $\boldsymbol{\mu_L}$ for a single electron via the relationship $\boldsymbol{\mu_L} = -(e/2m_e)\boldsymbol{L}$, where $e$ is the charge of the electron, $m_e$ is the mass of the electron.[1] In transition metal compounds, $\boldsymbol{L}$ is typically quenched by the crystal field i.e. $\boldsymbol{L} = \boldsymbol{0}$, and only emerges through the spin-orbit coupling (SOC) albeit much weaker than the crystal field interaction.[2] However, for 4$f$ ions such as $Tm^{3+}$, the crystal field interaction is less significant,[3] allowing $\boldsymbol{L}$ to contribute to the magnetic ground state.[2] Here, the significance of SOC becomes apparent, which takes the form $\lambda \boldsymbol{L}.\boldsymbol{S}$, where $\boldsymbol{S}$ is the spin angular momentum and $\boldsymbol{\lambda}$ is a constant. SOC can give rise to a variety of magnetic phenomena such as spin Hall effect, and topological magnetism.[4-8] These effects are decisive for the development of novel spintronic devices that require manipulation of the magnetization.

Recently, 4$f$-block magnetic insulators like thulium iron garnet ($Tm_3Fe_5O_{12}$, TmIG) were discovered to show enhanced perpendicular magnetic anisotropy (PMA).[9] This opens up exciting possibilities for promising applications, particularly in racetrack memory.[10-14] Interestingly, TmIG with a Curie temperature ($T_c$) of 560 K[15] and compensation temperature ($T_S$) of 85 K can be driven into different spin ordering states. Despite the expected quenching of $\boldsymbol{\mu_L}$ in Fe ions, it is likely to be significant in Tm. In this context, the study of $\boldsymbol{\mu_L}$ and spin magnetic moment ($\boldsymbol{\mu_S}$) contribution to enhanced PMA and magnetization dynamics in TmIG are scarce or almost non-existent. Previous investigations have focused primarily on the antiferromagnetic coupling between Tm and Fe, with Tm showing a significant temperature dependence.[16]

The techniques of ferromagnetic resonance spectroscopy (FMR) and X-ray magnetic circular dichroism (XMCD) are typically employed for precise measurement of orbital moments as they offer high signal-to-noise ratios.[17-18] Meanwhile, it is crucial to accurately determine $\mu_L$ in intrinsic TmIG without additional layers that might induce proximity effects while ensuring protection against surface oxidation. So far, XMCD studies on TmIG have either been with a Pt cap, which is known to introduce strong antisymmetric exchange and spin pumping effects at the interface[19-20] or uncapped exposed to surface degradation/oxidation.[12] Moreover, FMR studies on TmIG thin films are reported either on films made by sputtering[21-22] or under significant strain.[23]

Here, we investigate orbital and spin moments in a crystal environment of $\boldsymbol{f}$ and $\boldsymbol{d}$ ions in high-quality TmIG samples. Using the complementary techniques of FMR and XMCD spectroscopy, we explore the temperature dependence of these moments and establish a one-to-one comparison between the two techniques. While this combined approach has been employed in studies of other systems such as Co/Ni[24], YIG[25] and $Fe_3O_4$[26-27], they don't provide insights on $4f$ magnetism. We reveal that the uncharacteristically low value of the $\boldsymbol{g}$-factor in TmIG can be primarily attributed to a significant $\mu_L$ from $Tm^{3+}$ that opposes the dominant $\mu_S$ from $Fe^{3+}$. Finally, we discuss potential scenarios that account for minor differences in moment as determined from these two techniques.

## 2. Main Results

High-quality TmIG samples on SGGG substrates oriented along (111) were prepared by pulsed-laser deposition (PLD) technique (**Supplementary Note 1**). It is characterized by three individual magnetic sublattices, comprised of 3 $Tm^{3+}$ ions per formula unit (**f.u.**) on dodecahedral ($\boldsymbol{Ddh}$) sites, 2 $Fe^{3+}$/**f.u.** ions on octahedral ($\boldsymbol{Oh}$), and 3 $Fe^{3+}$/**f.u.** ions on tetrahedral ($\boldsymbol{Th}$) sites (**Figure 1a**). Atomic force microscopy (AFM) is used to examine the surface morphology of the films (Supplementary Note 1), and X-ray diffraction (XRD) method is performed to confirm the crystallinity of the TmIG samples (**Figure 1b**). TmIG film grown on SGGG (111) substrate is epitaxial, single-crystalline with orientation relationship <444> TmIG parallel to <444> SGGG. Increasing the film thickness causes a notable shift in the film peak, illustrating the strain propagation—from fully strained (12 nm) and partially structurally relaxed (20 nm) to fully relaxed (39 nm). The primary emphasis of our manuscript is on the 20 nm thickness, and comprehensive thickness-dependent results are outlined in Supplementary Note 2 and 5. **Figure 1c** shows a cross-sectional scanning transmission electron microscopy (STEM) image of ~10 nm TmIG coherently grown on SGGG with a sharp interface. High-magnified atomic arrangements in the plan-view image exhibit concentric hexagon patterns and confirm the state-of-the-art crystalline ordering through the whole film. Field-dependent normalized out-of-plane (OOP) and in-plane (IP) magnetic moments of TmIG samples at room temperature after subtracting the SGGG substrate's paramagnetic contribution are shown in **Figure 1d**. We observed a sharp magnetic reversal with a moment (~90 emu/cm$^3$) and small coercivity (3-3.5 Oe, expanded view in Supplementary Information) along the OOP magnetic field orientation. The small coercivity and sharp switching behaviour indicate PMA, which is expected for TmIG samples under tensile strain (TmIG/SGGG tensile strain ≈ 0.12%).[28] Saturation magnetization ($M_S$) of the TmIG samples increases as the temperature reduces consistent with previous TmIG samples[29] (Figure 1d, inset). The *M–T* curves can be fitted by the relationship $M = M_0(1 - T/T_C)^\beta$, where $\beta$ is the critical exponent, and ($M_0$) is the zero-temperature magnetic moment.[29] This power-law temperature dependence was attributed to dimensionality and surface effects in ultrathin magnetic films.[30-32]

The FMR measurements were carried out using a broadband co-planar waveguide (CPW) in a variable temperature cryostat under a maximum external OOP magnetic field ($H$) of 6 kOe and a frequency range from 2 to 26 GHz. All TmIG samples were placed film-upside-down (**Figure 2a**) onto the three-strip CPW. **Figure 2b** represents room-temperature FMR spectra for TmIG (20 nm)/SGGG sample at different microwave frequencies ranging from 5 to 15 GHz. The dimensionless dynamic parameters Gilbert damping $\alpha$ was calculated from Kittel equations (See **Figure 2c** and Supplementary Information). The temperature-dependent damping values show a monotonic decrease within the order of 10$^{-2}$, which is significantly larger than ~10$^{-5}$ in YIG.[33] The $\alpha$ values of the TmIG samples are comparable to previously reported values.[22-23, 32]

Under the FMR resonance condition, the microwave frequency, $\omega$ is proportional to the effective field, $H_{eff}$ as given by the Kittel equation $\hbar\omega = g_{eff}\mu_B\mu_0 H_{eff}$,[34] where $\hbar$ is the reduced Planck's constant, $\mu_0$ is the permeability of free space, and $\mu_B$ is the Bohr magneton. $g_{eff}$ can be represented approximately as $g_{eff} \cong 2 - C\lambda/\Delta$, where C is a constant of the order of unity, $\lambda$ is the SOC constant and $\Delta$ is an energy-level separation of different states [35-38]. Given that $\lambda$ of Fe$^{2+}$ = ~100 cm$^{-1}$ [39] and of Tm$^{3+}$ = ~2600 cm$^{-1}$,[40] Tm-based compounds are expected to have a much different $g_{eff}$. On the other hand, the reported $g_{eff}$ in transition metals are: bcc Fe ($g_{eff} = 2.09$), hcp Co ($g_{eff} = 2.18$), fcc Ni ($g_{eff} = 2.18$).[37] Using first-order perturbation theory, Kittel was able to relate $g_{eff}$ to the ratio of $\mu_L$ and $\mu_S$ [1, 34, 38, 41-42] as

$$g_{eff} = 2(1 + \mu_L/\mu_S) \quad (1)$$

Typically, a magnetic system with strong SOC has been shown to have non-zero $\mu_L$ resulting in $g_{eff} \neq 2$.[43-44] The $g_{eff}$ values for our TmIG samples are significantly lower than the value of 2 expected for a free electron and reduce monotonically from 1.7 at 330 K to 1.49 at 175 K (**Figure 2d**). The $g_{eff} < 2$ from our FMR data suggests that $\mu_L$ is strong and opposite in sign to $\mu_S$. In addition, the temperature-dependent data suggest that $\mu_L$ strengthens on lowering the temperature. It is to be noted that YIG, a close relative of TmIG, shows $g_{eff} = 2$. However, there are no *f* electrons in YIG, and hence, its SOC is expected to be negligible. Although $g_{eff}$ is also reported to be influenced by surface and interface effects that may lead to perturbation of the electron orbits,[24, 45] the observed reduced value cannot be entirely attributed to the dimensionality factor as thick TmIG samples also show $g_{eff} < 2$ (Supplementary Information). Our results suggest a critical role of Tm in influencing $g_{eff}$, which is inaccessible by FMR experiments. Therefore, a systematic element-specific investigation of $\mu_L$ and $\mu_S$ are further pursued in the next section.

Detailed element-specific X-ray magnetic circular dichroism (XMCD) spectroscopic measurements were performed at the BOREAS beamline[46] of the ALBA synchrotron to gain more insights into the temperature dependence of $g_{eff}$. Here, the absorption of circularly polarized X-ray photons ($\sigma^{\pm}$) resonant with core-level energies excites electrons with unbalanced spin, and the resulting imbalance of up and down spin in the unoccupied conduction band produces a large asymmetry in the absorption probability (**Figure 3a**, inset). All XMCD measurements were performed on TmIG samples that are capped with a protective layer of Ru (2 nm), which is expected to prevent surface degradation/oxidation without introducing any magnetic perturbations.[47] We particularly look at the spectra for the Fe $L_{2,3}$ (corresponding to a dipole-allowed $2p_{1/2}$ ($2p_{3/2}$) → $3d$ transitions), Tm $M_{4,5}$ ($3d → 4f$ transitions) and O $K$($1s → 2p$ transitions) edges, among all available elements of TmIG samples. The top panel of **Figure 3a-c** shows an exemplary X-ray absorption spectrum (XAS) at the Tm $M_{4,5}$, Fe $L_{2,3}$, and O $K$ edge for a TmIG samples (20 nm) recorded at 100 K and 300 K. The corresponding XMCD spectra, given by the difference between the absorption signal for $\sigma^+$ and $\sigma^-$ helicities, at these relevant energies are shown in Figure 3a-c, bottom panel. First, we look at the XAS and XMCD signals of Tm $M_{4,5}$ edge. The strong Tm $M_5$ XAS peak at ~1466 eV is attributed to $Tm^{3+}$ ion. Similar to our data, previous XMCD measurements also reported the shoulder peak at[13] ~1464 eV, which can be attributed to a small percentage of $Tm^{2+}$ ion.[48] We note a significant change in the strongest XMCD peak while the temperature varies from 300 K to 100 K. Therefore, the Tm XMCD signals are strongly temperature-dependent, consistent with previous literature[10] and as expected in most rare-earth (RE) iron garnets.[49] In the RE iron garnets, the magnetization of the RE sublattice can be regarded as that of a paramagnetic sublattice coupled to an effective exchange field caused by the non-compensated Fe magnetization and the neighbouring RE. Here, the magnetization of RE sublattice is given by the temperature-dependent Brillouin function of the effective exchange field.[50-51]

Next, we focus on XAS and XMCD peaks of Fe $L_3$ ($L_2$) edge with the maximum at ~709 eV (723 eV). The Fe **L**-edge is split into the $L_3$ and $L_2$ edges due to the spin-orbit coupling of the 2**p** core level. Consistent with previous reports on iron garnets,[10, 13, 16, 52] the XMCD spectra have two positive peaks from the $Fe^{3+}$ ions at the octahedral site (two per **f. u.**) and one large negative peak coming from the tetrahedral site (three per f.u.) of a TmIG unit cell, as a result of the antiferromagnetic coupling among the $Fe^{3+}$ ions (**Figure 3b**). The Fe $L_3$ energy peaks for the octahedral (*Oh*) and tetrahedral sites (*Td*) are similar to $Fe_3O_4$ thin films.[43, 53] In the bottom panel of Figure 3a-c, the XMCD spectra at 100 K and 300 K are compared. From the Fe $L_3$ energy peak, it is important to note that although the spectra from both *Oh* and *Td* sites are strong even at 300 K, a significant increase in signal is observed from the *Td* site (~66 %) and the *Oh* sites (~76 %) upon lowering the temperature.

Next, we examine the XAS and XMCD signals from the O **K** edge as they provide direct insights into the hybridization between the Fe 3**d** orbitals and the O 2**p** orbitals, as well as their magnetic properties, specifically their orbital moments. The XAS signal from the O **K** edge is reported in **Figure 3c** displays a double-peaked feature with a more intense peak at ~528.8 and a weaker peak at ~529.7 corresponding to the energy splitting of $e_g$ and $t_{2g}$ bands.[54] Interestingly, a clear non-zero dichroism with a low energy negative peak and a higher energy positive peak appears at the O **K** edge. This XMCD feature is similar to those from $Fe_3O_4$, where it was argued that the negative peak can be associated with the double exchange (DE) for the Fe(*Oh*)-O-Fe(*Oh*) via O 2*p* orbitals and the positive peak corresponds to the Fe(*Oh*)-O-Fe(*Td*) super-exchange (SE) interactions.[53, 55] Interestingly, we observe that the SE (positive XMCD peak) strengthens on lowering temperature, while the DE (negative XMCD peak) shows almost no temperature dependence.

## 3. Discussion

To facilitate a direct comparison of the $g_{eff}$ values obtained from the FMR results, we first calculate the total weighted $\mu_L$ and $\mu_S$, considering Tm and Fe elements present in the TmIG samples. **Figure 4a** and **b** show the temperature-dependent elemental values of $\mu_S$ and $\mu_L$ per f.u., respectively, obtained from sum-rule analysis of XAS and XMCD spectra. We utilize appropriate magneto-optical sum rules for Tm and Fe ions, which essentially represent weighted sums of the dichroic spectrum normalized by the spin-integrated x-ray absorption spectrum.[52] The sum rule equations for Fe $L_{2,3}$ ($2p \rightarrow 3d$) and Tm $M_{4,5}$ ($3d \rightarrow 4f$) absorption edges are described in Supplemental Material. Figure 4a and b display the spin and orbital moments, considering 5 Fe and 3 Tm ions per **f.u**. The Fe contribution to the spin moment, $\mu_S(\text{Fe})$, is found to be around 1.2 $\mu_B/\text{f.u.}$ at room temperature which is typically lower than the spin moment of bulk Fe due to the compensated Fe in the TmIG garnets.[56] These values are consistent with the literature values in other garnets. As seen in Figure 4a, Tm has a significantly weaker and opposite spin moment $\mu_S$ compared to Fe. Contributions to $|\mu_S|$ from both elements decrease with increasing temperature (Figure 4a). In contrast, the contribution of Fe ions to $\mu_L$ is almost negligible and exhibits a weak temperature dependence, whereas Tm ions demonstrate a larger $|\mu_L|$, larger than their spin counterpart, which is characterized by a strong temperature variation, as depicted in **Figure 4b**.

Our previous moment analysis in Figure 4a and b implies that $\mu_L$ is predominantly from Tm and $\mu_S$ is predominantly from Fe contribution. We compare $\mu_L/\mu_S$ ratio per f.u. as a function of the temperature using net $\mu_L/\mu_S$ (open symbols) and $\mu_L(\text{Tm})/\mu_S(\text{Fe})$ (solid symbols) as shown in **Figure 4c**. Again, the net $\mu_L/\mu_S$ (open symbols) is determined by considering 5 Fe and 3 Tm ions per f.u. Indeed, we find that the curves trace each other for $T >$ 150 K. With this high-temperature approximation, **Equation (1)** simplifies to

$$g_{eff} \approx 2\left(1 + \frac{\mu_L(\text{Tm})}{\mu_S(\text{Fe})}\right) \quad (2)$$

To check the validity of this approximation, we explicitly calculated the total orbital and spin moments (as $g_{Total}$ from Equation 1) and compared it with the simplified assumption using site-specific **Equation (2)** (as $g_{Simplified}$). **Figure 4d** shows the comparison of $g_{eff}$ obtained from FMR ($g_{FMR}$) with that from XMCD using Equation (1) ($g_{Total}$) and (2) ($g_{Simplified}$). There is also a good qualitative agreement between FMR and XMCD experimental values of $g$-factors. Hence, we argue that the reduction of $g$-factor in TmIG as determined from FMR can be primarily explained by the relative contribution of spin and orbital moments. In particular, following Equation (1) and (2), the reduction of $g_{eff}$ can be attributed to the significant orbital moment contribution of Tm $\mu_L$(Tm) that is the opposite in sign to $\mu_S$ (Fe). Equation (2) also explains the temperature dependence of $g_{eff}$ given the strong modulation of $\mu_L$(Tm) with temperature.

We note that FMR gives a slightly larger $g_{eff}$ than XMCD and the deviation increases on lowering the temperature. For a TmIG (20 nm) samples, FMR yields $g_{FMR} \approx 1.7$ while XMCD yield $\mu_L/\mu_S = -0.25$ and $g_{Total} = $ ~1.5 at room temperature. It was recently shown by Shaw et al.[57] in metallic systems that quantitative discrepancies in $\mu_L/\mu_S$ determined from XMCD and FMR can be explained by invoking the second order spin mixing correction ($b^2$) to Equation (1). However, the inclusion of second-order spin-mixing parameters determines a non-negative correction to Equation (1), which requires FMR to underestimate the moment values opposite to the present case. In addition, it was recently reported that spin mixing correction is not required in the closely similar YIG system.[25] We speculate that the discrepancy could arise from one or both of the following scenarios - (i) uncertainty in accurate determination of Tm moments (Supplementary Information), (ii) correction in Equation (1) when considering ferromagnetic oxides with 4$f$ ions with large orbital moment. We rule out the potential contribution from oxygen via hybridization with Fe $d$ orbitals as negligible. This conclusion is supported by the investigation of YIG, which has the same Fe sub-lattices, revealing insignificant spin-mixing due to the low SOC in $Fe^{3+}$.[25] In addition, we find that this discrepancy cannot be explained by simple estimation of antiferromagnetic sub-lattices where one sublattice is strongly damped (Supplementary Information). It is worth noting that the discrepancy in $g_{eff}$ widen on decreasing temperature. Coincidently, the strength of $\mu_L$ contribution from Tm, i.e., $|\mu_L(\text{Tm})|$ increases with lowering temperature (Figure 4b) suggesting scenario (ii) maybe the more influential factor in explaining the phenomenon. Notably, another parameter exhibiting significant temperature dependence is $\mu_S$(Fe), which also increases upon temperature reduction.

## 4. Conclusion

We investigated orbital and spin moments in a high-quality TmIG oxide magnetic insulator grown on SGGG substrates using complementary FMR and XMCD spectroscopy techniques with qualitative agreement. Systematic sum-rules analysis of the temperature-dependent XMCD reveals a significant disparity in the orbital-to-spin moment ratio between Tm and Fe ions. Our findings suggest that the unusually low value of $g$-factor in this system can be mainly attributed to a large orbital moment of $Tm^{3+}$ ion, which opposes the dominant spin moments from $Fe^{3+}$ ions. Our study presents a model system for understanding orbital moments in an environment of *d* and *f* ions in high-quality films. The possibility to modify the *g*-factor by doping or substituting with *f* ions is especially appealing for fundamental studies on probing local magnetic environment by just using long-range magnetic spectroscopy techniques.

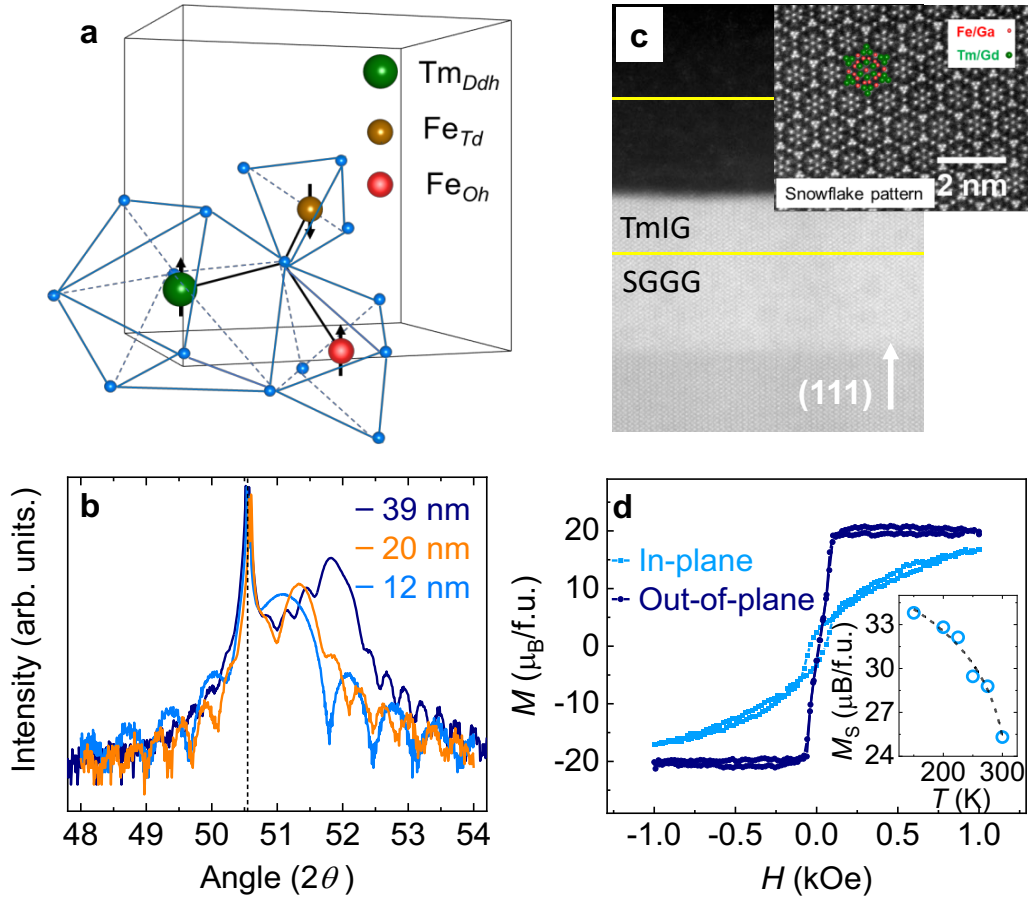

**Figure 1. Fundamental characteristics of Tm$_3$Fe$_5$O$_{12}$ thin films.** (a) Crystal structure of Tm$_3$Fe$_5$O$_{12}$, representing Fe in its octahedral ($Oh$), tetrahedral ($Td$), and Tm in dodecahedral ($Ddh$) positions. (b) X-ray diffraction scans of TmIG (12, 20 and 39 nm) grown on (111)-oriented SGGG substrates indicating fully strained, partially relaxed and fully relaxed TmIG layer, respectively. The vertical dotted line is the substrate peak. (c) Cross-sectional STEM image for a TmIG/SGGG (111) sample viewed along the [1$\bar{1}$0] direction. The yellow line marks the boundary between surface, TmIG and SGGG layers. High-magnification plan-view STEM image with an overlay of the [111]-projection of the TmIG lattice to highlight the snowflake pattern. (d) OOP and IP magnetic hysteresis loops after subtracting the paramagnetic substrate background (Saturation magnetization as a function of temperature, inset). The dashed lines represent power-law fits.

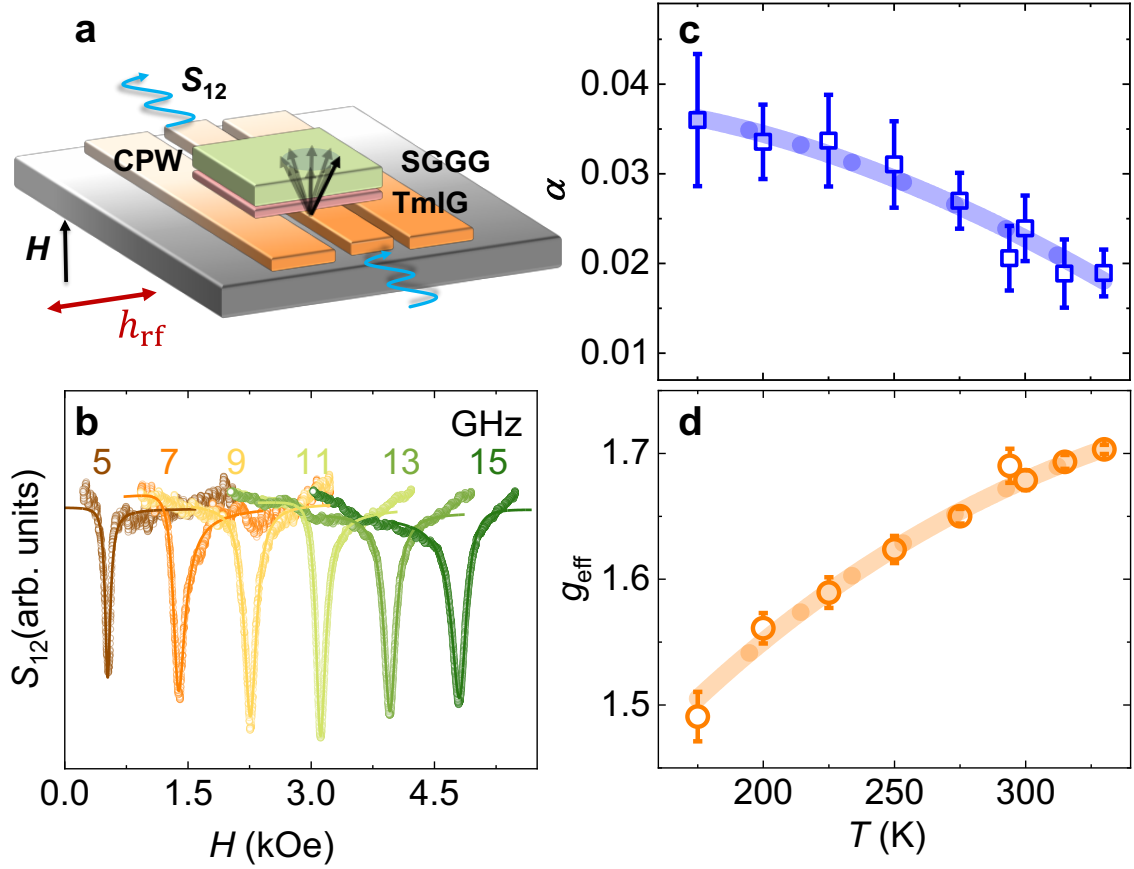

**Figure 2. Ferromagnetic resonance (FMR) spectroscopy results.** (a) Schematic diagram of a FMR spectroscopy with co-planar waveguide (CPW) in an OOP magnetic field configuration. (b) FMR spectra at various frequencies in OOP field sweep configuration for TmIG(20 nm)/SGGG at room temperature. Temperature-dependent (c) $\alpha$ and (d) $g_{\text{eff}}$ from OOP FMR measurements.

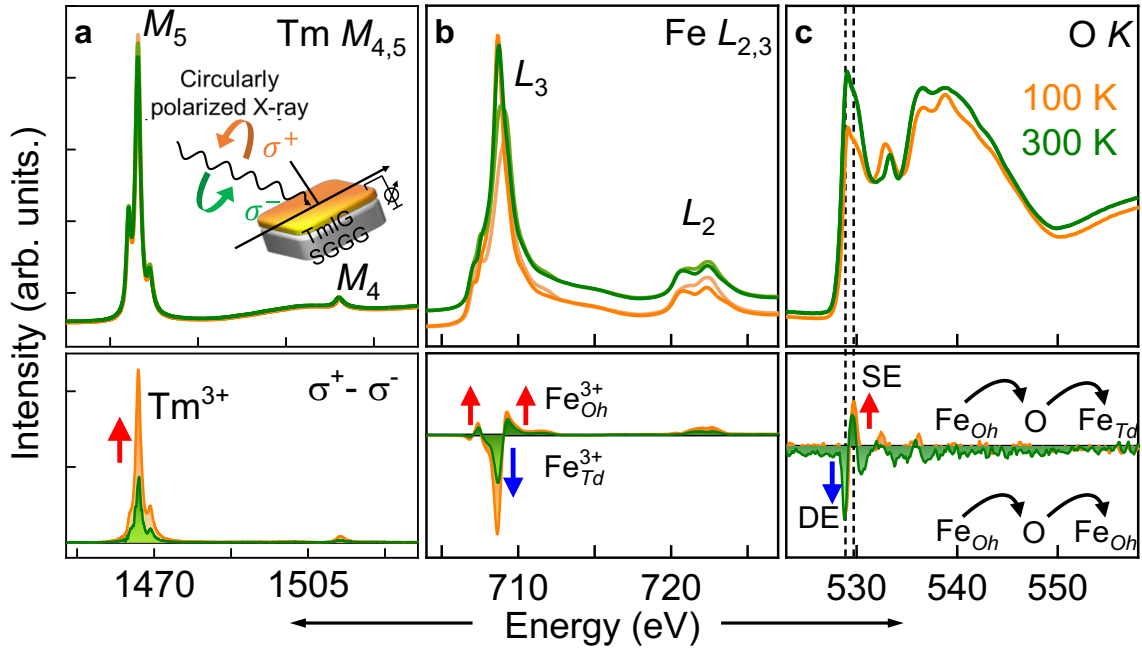

**Figure 3. X-ray magnetic circular dichroism (XMCD) results.** (a) Top and bottom panel shows XAS and XMCD intensity spectra of the Tm $M_{4,5}$ edge recorded at 100 K and 300 K, respectively. Inset showing XMCD experimental geometry with circularly polarized X-ray. (b) XAS (top) and XMCD (bottom) intensity spectra of the Fe $L_{2,3}$ edge recorded at 100 K (orange) and 300 K (green). The arrows indicate the relative orientation of magnetic moments (at octahedral and tetrahedral positions). (c) XAS (top) and XMCD (bottom) intensity spectra of the O $K$-edge recorded at 100 K and 300 K. All measurements are performed at out-of-plane field configuration in a 1 T field. All plots are rescaled in arb. units.

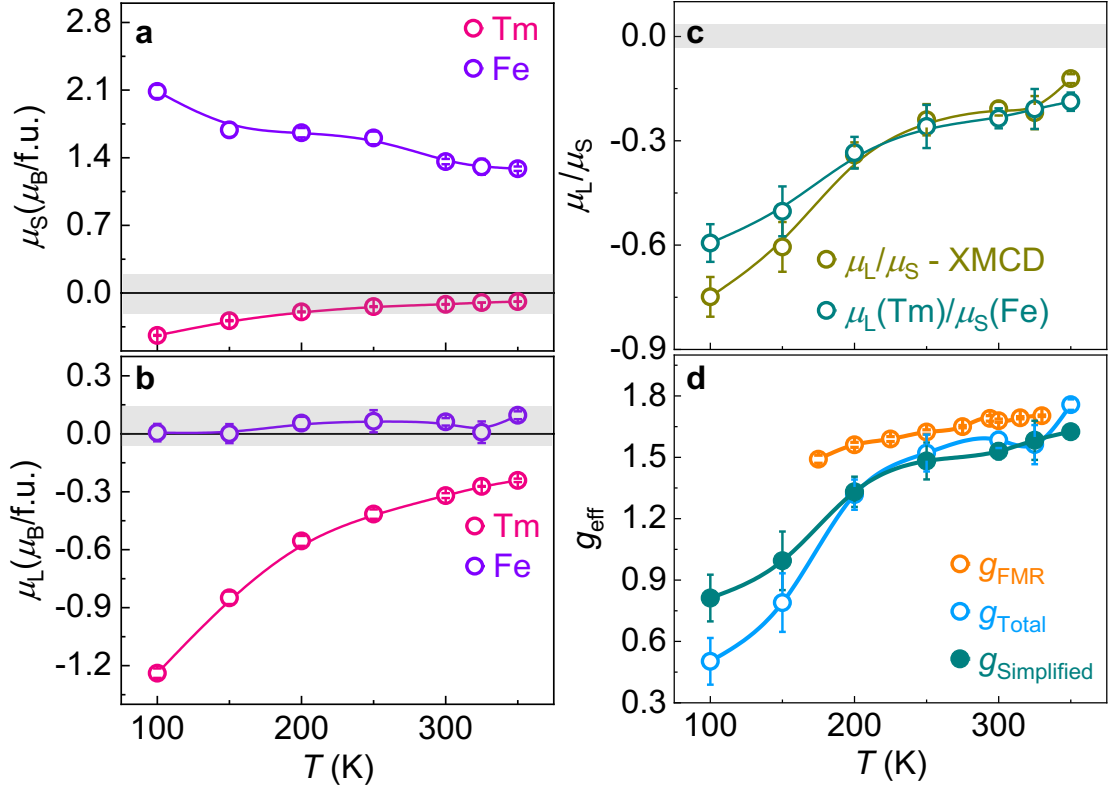

**Figure 4. *g*-factor analysis from XMCD and FMR spectroscopy.** (a) Orbital moment ($\mu_L$) and (b) spin moment ($\mu_S$) per f.u. as a function of temperature for Tm (dark pink) and Fe (violet) as determined from XMCD measurements at normal incidence. (c) Ratio of $\mu_L$ to $\mu_S$ as a function of temperature calculated for net ($\mu_L/\mu_S$) and ($\mu_L$(Tm)$/\mu_S$(Fe)). (d) $g_{\text{eff}}$ calculated from FMR ($g_{\text{FMR}}$, XMCD using equation (1): $g_{\text{Total}}$, and simplified equation (2): $g_{\text{Simplified}}$.

**Author Contributions**

G.J.O, J.L., and A.A. designed research; G.J.O. synthesized the samples, performed fundamental characteristics and magnetometry. P.G. and M.V. performed XMCD measurements; J.L. and A.G. performed FMR measurements; G.J.O., J.L., P.G. and M.V. analyzed data; A.A., J.L., and G.J.O. interpreted results; and Z.S.L., S.P., T.S.S., S.T.L. helped to write paper.

**Acknowledgements**

This research is supported by the Agency for Science, Technology and Research (A*STAR) under its Advanced Manufacturing and Engineering (AME) Individual Research Grant (IRG) (A2083c0054) and A*STAR SpOT-LITE programme (A*STAR Grant No. A18A6b0057) through RIE2020 funds from Singapore. G.J.O, Z.S.L. and A.A. acknowledge the National Research Foundation (NRF) of Singapore under its NRF-ISF joint program (Grant No. NRF2020-NRFISF004-3518) for the financial support. We acknowledge beamtime at ALBA synchrotron BL29 through proposal #2021085233. We would also like to acknowledge Xiao Chi for his technical suggestions on XMCD.

**Conflict of Interest Statement**

The authors declare no competing interests.

**Data Availability Statement**

The data that support the findings of this study are available from the corresponding author upon reasonable request.

Supporting Information

**Room temperature strong orbital moments in perpendicularly magnetized magnetic insulator**

Ganesh Ji Omar, Pierluigi Gargiani, Manuel Valvidares, Zhi Shiuh Lim, Saurav Prakash, T. S. Suraj, Abhijit Ghosh, Sze Ter Lim, James Lourembam[*], Ariando Ariando[*]

Ganesh Ji Omar, Saurav Prakash, T. S. Suraj, Ariando Ariando
Department of Physics, National University of Singapore, Singapore 117542, Singapore
Email: ariando@nus.edu.sg

Pierluigi Gargiani, Manuel Valvidares
CELLS-ALBA Synchrotron Radiation Facility, Cerdanyola del Valles, Spain

Zhi Shiuh Lim, Abhijit Ghosh, Sze Ter Lim, James Lourembam
Institute of Materials Research and Engineering, A*STAR (Agency for Science, Technology and Research), 2 Fusionopolis Way, Innovis, Singapore 138364, Singapore
Email: james_lourembam@imre.a-star.edu.sg

Keywords: Magnetic Insulators, Orbital moments, Magnetic Spectroscopy, Electronic *g*-factor

This PDF file includes:
Supplementary Notes 1-6
Supplementary Figures S1-S8
Supplementary Table S1

**Supplementary Note 1: High-quality TmIG thin-film growth and quality characterization**

Single crystalline epitaxial thulium iron garnet ($Tm_3Fe_5O_{12}$, TmIG) samples are synthesized on lattice-matched substituted-GGG (SGGG, 12.381 Å) substrates along (111) crystallographic orientation. Each sample is prepared and deposited using a commercial TmIG polycrystalline target (purity 99.99%). Prior to the deposition of films, the target surface is pre-ablated and rastered using a few thousand pulsed laser shots. Optimal growth parameters are summarized in the table below. After deposition, the sample was cooled down to room temperature at a relatively slower rate (10°C/min) under the same $pO_2$ pressure as during the growth.

| Laser Repetition Rate | 10 Hz |
|---|---|
| Laser Energy Density | ~ 1.5 J/cm$^2$ |
| Substrate Temperature | 600°C |
| Oxygen Partial Pressure ($pO_2$) | 200 mTorr |
| Target | polycrystalline TmIG |

**Table S1** | Pulse laser deposition growth parameters for TmIG film on SGGG substrate.

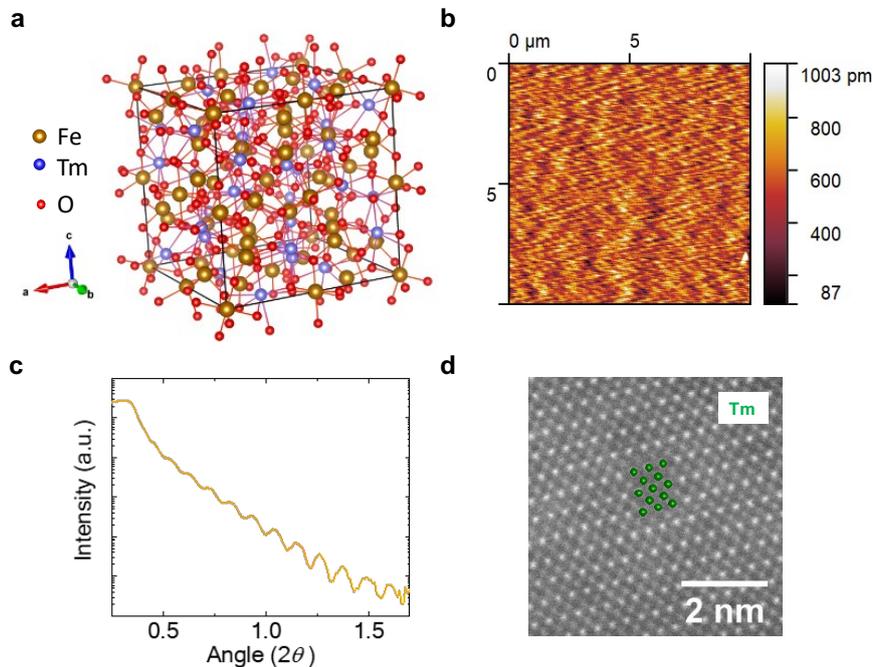

**Figure S2** | (a) Schematic illustration of one unit cell of TmIG crystal structure, consisting of 8 sub-lattices. (b) Atomic Force Microscopy (AFM) topography image, (c) Kiessig fringes in X-ray Reflection (XRR) of a TmIG/SGGG sample. (d) Cross-sectional scanning transmission electron microscopy (STEM) image. The green dots denote the locations of Tm in TmIG crystal.

$Tm_3Fe_5O_{12}$ (TmIG) garnet has a cubic unit cell. Within this unit cell, there are 8 sub-lattices. One sub-lattice corresponds to one formula unit. In one unit cell, there are 24 $Tm^{3+}$ dodecahedral atomic sites (blue) with 8 oxygen (red) neighbours, 16 $Fe^{3+}$ tetrahedral atomic sites (brown) with 6 oxygen neighbours, 24 $Fe^{3+}$ octahedral atomic sites (brown) with 4 oxygen neighbours. The schematic illustration for one unit cell is given in Figure S1(a). Atomic Force Microscopy (AFM) and X-ray reflection (XRR) methods are used to observe the structural and morphological characteristics of a TmIG sample, as shown in Figure S1(b) and (c), respectively. AFM topography image and Kiessig fringes of the TmIG sample (39 nm) show that the film surface root-mean-square roughness is 1.3 pm, where pm is picometers. Additionally, the crystal quality structure was examined using high-resolution scanning transmission electron microscopy (STEM) of a cross-section of a TmIG/SGGG sample [Figure S1(d)]. We calculated the dead layer in our samples and found to be ~ 2-3 nm.

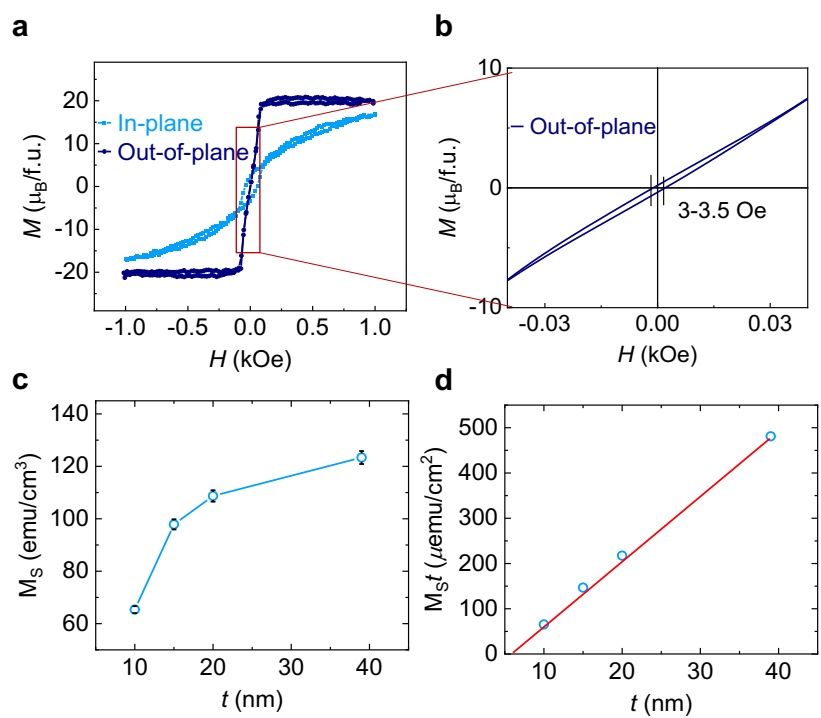

**Figure S2** | (a) OOP and IP magnetic hysteresis loops after subtracting the paramagnetic substrate background. (b) Expanded view showing small coercivity around 3-3.5 Oe. (c) Saturation Magnetization as a function of TmIG sample thickness $t$ (nm). (d) Areal magnetic moment as a function of TmIG thickness, which indicates a negligible magnetic dead layer.

In addition to the dead layer calculation, we have also explored the magnetoelastic contribution to the magnetic free energy for films with different thicknesses. The magnetoelastic contribution to the magnetic free energy ($E_{ME}$) along the [111] direction is given by $E_{ME} = 3\lambda_s C_{111} \epsilon \sin^2\theta = K_{ME} \sin^2\theta$, where $K_{ME}$ is the magnetoelastic energy, $\lambda_s$ is the saturation magnetostriction along the $\langle 111 \rangle$ directions, $C_{111}$ is the elastic constant, $\epsilon$ is the strain along the out-of-plane [111] axis, and $\theta$ is the angle between the magnetization and the strain axis. From the literature, we obtain $\lambda_s = -5.2 \times 10^{-5}$ and $C_{111} = 7 \times 10^{-6}$ dyn/cm² for TmIG[1]. Based on our XRD data, we measure an out-of-plane strain of -1.09 % in the strained films, leading to a magnetoelastic energy of $\sim 1.3 \times 10^5$ erg/cm³. We have plotted the magnetoelastic energy across the three films of different thickness, representing the transition from fully strained to fully relaxed states (Figure S3).

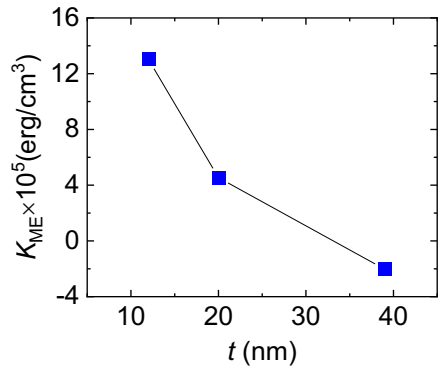

**Figure S3** | Magnetoelastic energy as a function of TmIG sample thickness *t* (nm).

**Supplementary Note 2: Discussion on magnetodynamics experiments using Ferromagnetic Resonance (FMR)**

In our ferromagnetic resonance spectroscopy (FMR) experiments, the change in the transmittance, $S_{12}$ is a field-dependent susceptibility of the TmIG layer can be represented as a combined Lorentz absorption. Hence, we have used the equation below to fit absorption peaks at the resonant frequency.

$$S_{12} = \frac{A \Delta H^2}{[(H - H_{res})^2 + \Delta H^2]} + \frac{B \Delta H (H - H_{res})}{[(H - H_{res})^2 + \Delta H^2]} + C \qquad (1)$$

where $\Delta H$ is the linewidth of the spectrum or full width of half maximum (FWHM), $H_{res}$ is the peak resonance field, $A$ and $B$ are Lorentzian coefficients, and C represents the spectral background.

The linear relation between the microwave frequencies ($f$) and resonant fields ($H_{res}$) is fitted using the Kittel equation in the resonant condition with the field in the OOP configuration as described below,

$$f = \gamma_0/2\pi \, (H_{res} - \mu_0 M_{eff}) \qquad (2)$$

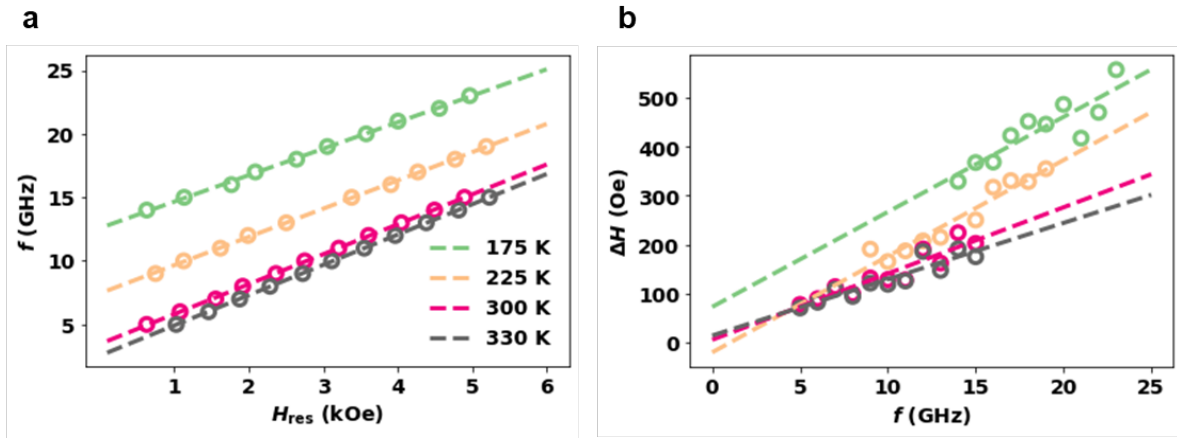

**Figure S4** | (a) Temperature-dependent frequency ($f$) vs resonance field ($H_{res}$). Dashed lines are fitting to Kittel equation, (b) Temperature-dependent FMR linewidth ($\Delta H$) as a function of frequency ($f$) for a TmIG(20 nm)/SGGG sample.

Effective magnetization ($M_{eff}$) is determined from FMR experiments. We have extracted $\gamma_0$, gyromagnetic ratio parameter by fitting the plots above using Equation 2. Subsequently, damping, $\alpha$ has been estimated from the slope using:

$$\Delta H = \frac{4\pi\alpha}{\gamma_0}f + \Delta H_0 \quad (3)$$

Here $\Delta H_0$ represents the inhomogeneous linewidth broadening. It is a frequency-independent linewidth contribution arising from inhomogeneities in magnets. The spectroscopic splitting factor, $g_{eff}$ is calculated from $\gamma_0 = g_{eff}\mu_B/\hbar$ relation. The plots, along with the fits corresponding to equations (2) and (3) for a 20 nm TmIG/SGGG are shown in Figure S4(a) and (b), respectively.

Furthermore, $M_{eff}$ as determined from equation (2) shows strong temperature dependence. Negative values of $M_{eff}$ indicate perpendicular magnetic anisotropy (PMA), which gets stronger on lowering temperature[2]. Additional FMR measurements in OOP configuration were carried out in other TmIG film thicknesses. Figure S5(b) and (c) shows thickness-dependent results of $\alpha$ and $g_{eff}$, respectively taken at room temperature. While $\alpha$ shows thickness dependence, $g_{eff}$ remains constant around 1.67 over a wide range of thickness values.

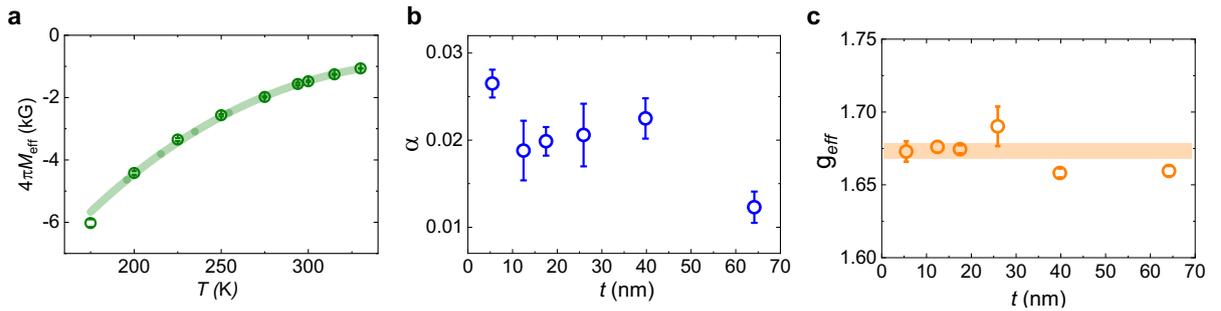

**Figure S5** | (a) Temperature dependent effective magnetic saturation ($M_{eff}$) of TmIG (20 nm)/SGGG sample, TmIG sample thickness ($t$ nm) dependent (b) Gilbert damping parameter, $\alpha$ and (c) spectroscopic splitting factor, $g_{eff}$.

**Supplementary Note 3: XAS/XMCD measurement technique**

The XAS/XMCD experiments were performed at the BOREAS beamline in the high-field cryo-magnet end station HECTOR of the ALBA synchrotron [3]. The signal was collected in total-electron-yield mode (TEY) by measuring the sample drain current to the ground using a Keithley 428 current amplifier. The spectra were normalized by the impinging X-ray flux measured as the drain current signal on a freshly evaporated gold mesh placed after the last optical element of the beamline. The XMCD measurements were performed by switching the impinging X-rays helicity, produced by an APPLE-II type undulator, while keeping an applied magnetic field of 1 T collinear to the impinging X-ray beam and perpendicular to the sample surface. The spectra at Fe $L_{2,3}$ edge energy was collected with a degree of circular polarization higher than 99%, while the spectra at the Tm $M_{4,5}$ were collected with a circular polarization degree >90%. The sample temperature was determined via the temperature of the cryostat cold finger.

**Supplementary Note 4: Discussion on Sum rule analysis**

The XMCD sum rules [4,5] were employed to determine the spin and orbital moment of the Tm and Fe. The spectra were normalized by a multiplicative factor, determined by averaging the spectra over a determined interval on the pre-edge region to correct for eventual signal drifts. The XMCD and XAS signals were respectively determined as the difference and sum between spectra collected with opposite helicities. The XMCD signal collected at the Tm $M_{4,5}$ edges were further multiplied by a factor of 1/0.9 to account for the lower than 1 polarization degree of the impinging X-rays. In order to apply the sum rules to Fe (Tm), three different integrals must be calculated, namely: $p$ the integral of the XMCD intensity over the $L_3(M_5)$ absorption edge, $q$ the integral of the XMCD intensity over the full spectrum and $r$ the integral of the XAS signal over the full spectrum after subtracting the contribution to the XAS intensity determined by the transition to the continuum states. The latter was estimated for Fe as a double-step function [6], while for Tm $M_{4,5}$, due to the shape of the background and the intensity of the $M_{4,5}$ transition, it was estimated using an asymmetrically reweighted penalized least squares smoothing function [7]. The spin and orbital moment operator's expectation values were determined for the Fe ion as:

$$\langle S_{\text{eff}}\rangle_{\text{Fe}} = 2\langle S_z\rangle + 7\langle T_z\rangle = \left(\frac{6p - 4q}{r}\right)N_h, \langle L_z\rangle = \left(\frac{4q}{3r}\right)N_h \qquad (4)$$

corresponding to the $2p \rightarrow 3d$ dipole transitions and for the Tm ion as:

$$\langle S_{\text{eff}}\rangle_{\text{Tm}} = 2\langle S_z\rangle + 6\langle T_z\rangle = \left(\frac{5p - 3q}{r}\right)N_h, \langle L_z\rangle = \left(\frac{2q}{r}\right)N_h \qquad (5)$$

corresponding to the $3d \rightarrow 4f$ transitions. Here, $N_h$ is the number of holes in the final $3d$ ($4f$) shell for Fe (Tm) and $T_z$ is the z-projection of the spin-dipole operator. $r$ refers to the integrated value of the total XAS (after subtracting the step function) at the $L_2$ and $M_4$ post-edges, as shown in Figure S6(a) and (b). The parameters $p$ and $q$ refer to the integrated values of the XMCD at $L_3$ ($M_5$) and $L_2$ ($M_4$) post-edges, respectively (see Figures S6(c) and (d)). $\langle L_z\rangle$, $\langle S_z\rangle$, $\langle T_z\rangle$ are the expectation values with respect to the Z component of the orbital angular momentum, the spin angular momentum, and the magnetic dipole operator of the $3d$ ($4f$) shell for $L_{2,3}$ ($M_{4,5}$) edges, respectively. Error bars on the spin and orbital moment obtained via the sum rules analysis applied to the experimental spectra have been estimated by varying the $p$, $q$ and $r$ parameters integration limits within a reasonable interval. It has been assumed that the errors on $\mu_S$ and $\mu_L$ arising from the incorrect estimation of the number of holes and spin-orbit mixing of valence states in $\mu_S$ evaluation are smaller compared to the ones derived from $p$, $q$ and $r$ integral estimations.

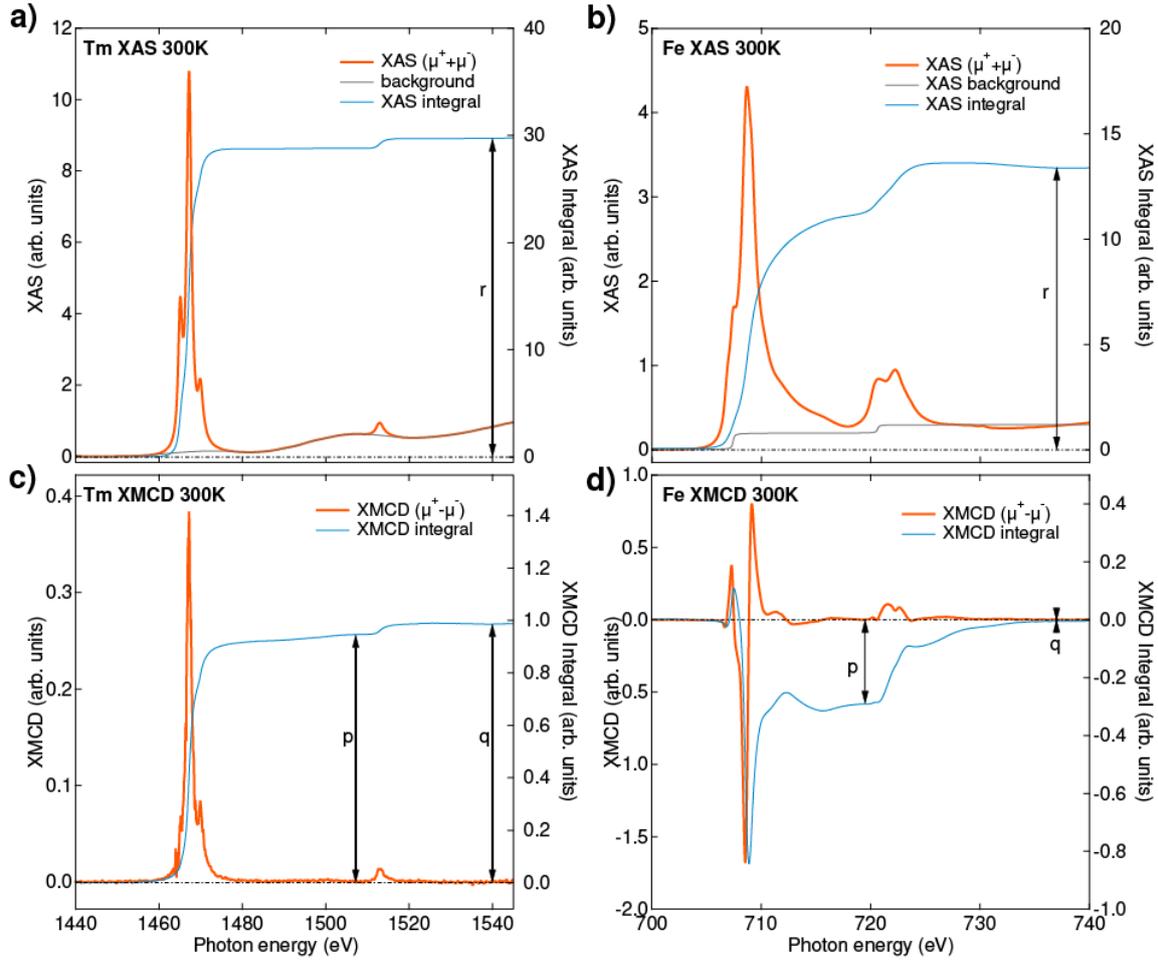

**Figure S6** | XAS spectra, total XAS integral and background subtraction of (a) Tm edge (b) Fe edge at 300 K of TmIG film. XMCD spectra and total XMCD integral of (c) Tm edge (d) Fe edge at 300 K of TmIG film.

For Fe, we assume that the $T_z$ value, due to the cubic symmetry environment of the Fe ion, to be negligible, allowing us to determine the expectation value of the $Z$ projection of the spin operator directly from the spin sum rules. This assumption does not hold for the rare earths in general and in particular for the Tm ions. In this case, we have assumed that the ratio $\langle T_z \rangle_{\text{free}} / \langle S_z \rangle_{\text{free}}$ calculated on the free Tm ion [8,9] holds for the studied system and is independent of temperature. This assumption holds true if the perturbation of the crystal field is smaller than the spin-orbit coupling, a condition normally satisfied for lanthanide elements [10]. If we include apply this assumption to equation 5, the spin sum rule takes the form:

$$\langle S_{\text{eff}} \rangle_{\text{Tm}} = \langle S_z \rangle \left( 2 + 6 \frac{\langle T_z \rangle_{\text{free}}}{\langle S_z \rangle_{\text{free}}} \right) = \left( \frac{5p - 3q}{r} \right) N_h \qquad (6)$$

$$\langle S_z \rangle = \frac{\left( \frac{5p - 3q}{r} \right) N_h}{\left( 2 + 6 \frac{\langle T_z \rangle_{\text{free}}}{\langle S_z \rangle_{\text{free}}} \right)} \qquad (7)$$

The values of $\langle T_z \rangle_{\text{free}}$ and $\langle S_z \rangle_{\text{free}}$ for $Tm^{3+}$ have been taken from Table I in ref [8]. A further source of uncertainty for the application of the spin sum rule to the lanthanide XMCD spectra comes from the possible mixing of the $3d_{5/2}$ and $3d_{3/2}$ states due to $3d - 4f$ hybridization [8,11]. The hybridization of the $3d_{5/2}$ and $3d_{3/2}$ states will mix the spectral weight between the $M_4$ and $M_5$ edges thus making inapplicable the spin sum rules even in presence of a large energy separation of the two absorption edges. However, it has been demonstrated using atomic full multiplet calculations that the mixing is particularly large only for less than half filled 4f orbitals, while it is below 10% for more than half filled ones [11]. In particular for $Tm^{3+}$ ions the $3d_{5/2}$ and $3d_{3/2}$ mixing is of the order of 1%, thus making the spin sum rule applicable. Since we are calculating moments for a unit cell, the number of holes per unit cell we have assumed $N_h$ = 4.7×5 for the Fe ion[12] and =2×3 for the Tm ion according to the nominal 3+ oxidation state.

**Supplementary Note 5: Thickness dependent XMCD on TmIG**

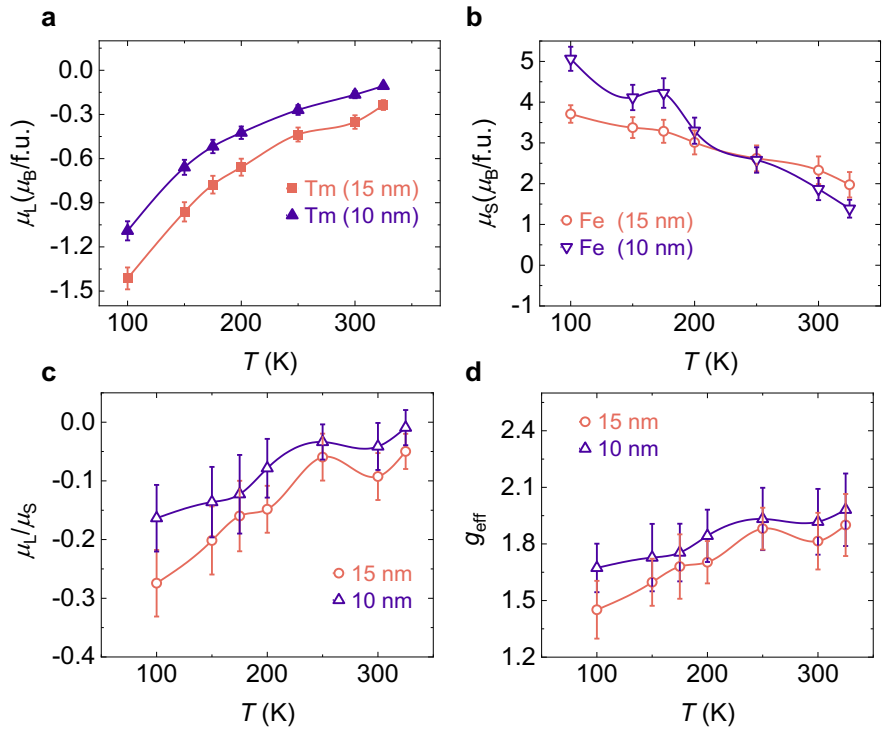

**Figure S7** | (a) Orbital moment ($\mu_L$) and (b) spin moment ($\mu_S$) per formula unit as a function of temperature for Tm and Fe measured at 1 T field shown for two different TmIG thicknesses: 10 nm (purple) and 15 nm (orange). Values of (c) $\mu_L/\mu_S$ and (d) $g_{eff}$ calculated from the XMCD.

Additional XMCD measurements were carried out on 10 nm and 15 nm thick TmIG films, as shown in Figure S7. Similar to the 20 nm thick films, we observe that $|\mu_L(\text{Tm})|$ is significant and increases with lowering temperature at these thicknesses, too. Furthermore, $|\mu_L(\text{Tm})|$ had the strongest variation in temperature compared to other spin and orbital moments. The lower values of $|\mu_L(\text{Tm})|$ for the thinner 10 nm sample can be ascribed to the contribution of the 2-3 nm magnetic dead layer to the XAS intensity, resulting in a reduced value of the XMCD-derived estimation of $L$ and $S$ [Figure S7(a)]. This resulted in a smaller $|\mu_L/\mu_S|$ in thinner films, as shown in Figure S7(c). Consequently, we find that the calculated value of $g_{\text{eff}}$ is higher in thinner films [Figure S7(d)]. This thickness-dependent $g_{\text{eff}}$ is markedly different from the FMR measurements, where we find a thickness-independent trend and adds another discussion point on the contrast between these two techniques. However, we must note that the XMCD-derived $g_{\text{eff}}$ is inherently affected by larger experimental error thus its thickness dependence cannot be entirely considered robust or outside the experimental error.

**Supplementary Note 6: Discussion on $g$-factor in ferrimagnetic systems**

The TmIG structure is composed of three sublattices at crystallographic: $Tm^{3+}$ occupying dodecahedral (*Ddh* site), and $Fe^{3+}$ occupying octahedral (*Oh* site) and tetrahedral (*Td* site) sites[13]. According to the theory of antiferromagnetic resonance in systems with sub-lattices, the resonance parameter can be represented by magnetizations of the individual sub-lattices[14]. By simplifying the Fe sites to be of one sub-lattice and under the assumption of large damping from the $Tm^3$ ion, the effective spectroscopic $g$-factor, $g_{\text{eff}}$ is given by[15]

$$g_{\text{eff}} = \frac{g_{\text{Fe}}(\mu_{\text{Fe}} - \mu_{\text{Tm}})}{\mu_{\text{Fe}}} \approx \frac{2(\mu_{\text{Fe}} - \mu_{\text{Tm}})}{\mu_{\text{Fe}}} \qquad (8)$$

where $\mu_{\text{Fe}}$ and $\mu_{\text{Tm}}$ are the magnetizations at the Fe and Tm sub-lattices, respectively. Here, $g_{\text{Fe}} \approx 2$ since SOC is weak, Fe sublattice is weak. To calculate the total moment, we use $\mu_z = -\mu_B L_Z - 2\mu_B S_Z$. Figure S7(a) shows the $\mu_z$ values plotted against temperature for Tm, Fe and total. To check whether equation (8) holds true in our case, we use the values of $\mu_z$ and calculate $g_{\text{eff}}$, which we denote as $g_{\mu Z}$. This is compared with those from FMR: $g_{\text{FMR}}$ and from Kittel equation: $g_{\text{XMCD}}$ as shown in Figure S8(b). Although $g_{\mu Z}$ follows similar temperature dependence, we find that quantitively, it has a larger deviation from $g_{\text{FMR}}$.

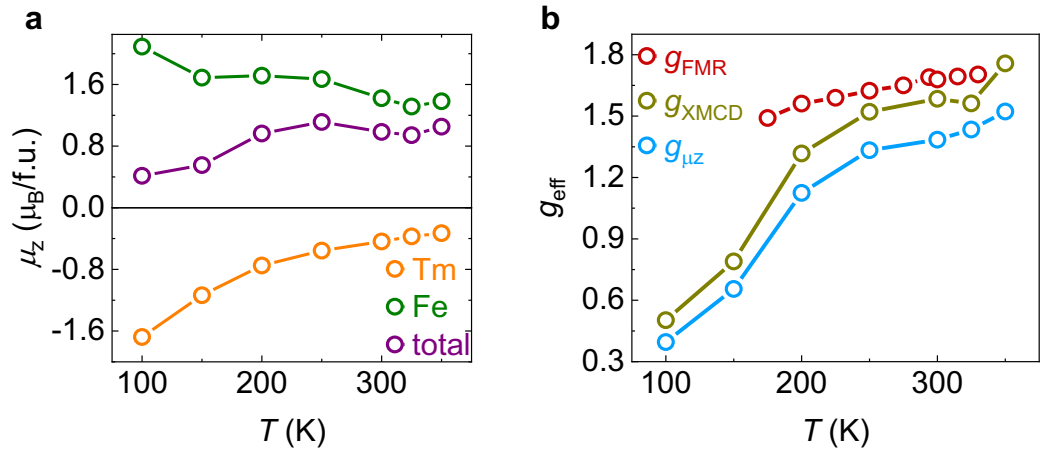

**Figure S8** | (a) Total magnetic moment ($\mu_z$) per formula unit as a function of temperature for Tm, Fe and including both at 1 T field. (b) $g_{eff}$ calculated from FMR, XMCD and Total moment ($\mu_z$) as a function of temperature for TmIG.

**Supplementary References:**

[S1]  A. Paoletti, *Physics of magnetic garnets*, Vol. 70, Elsevier Science & Technology, **1978**.

[S2]  J. M. Shaw, H. T. Nembach, and T. J. Silva, Applied Physics Letters **105** (2014).

[S3]  A. Barla *et al.*, Journal of Synchrotron Radiation **23**, 1507 (2016).

[S4]  B. T. Thole, P. Carra, F. Sette, and G. van der Laan, Physical Review Letters **68**, 1943 (1992).

[S5]  P. Carra, B. T. Thole, M. Altarelli, and X. Wang, Physical Review Letters **70**, 694 (1993).

[S6]  C. T. Chen, Y. U. Idzerda, H. J. Lin, N. V. Smith, G. Meigs, E. Chaban, G. H. Ho, E. Pellegrin, and F. Sette, Physical Review Letters **75**, 152 (1995).

[S7]  S.-J. Baek, A. Park, Y.-J. Ahn, and J. Choo, Analyst **140**, 250 (2015).

[S8]  Y. Teramura, A. Tanaka, B. T. Thole, and T. Jo, Journal of the Physical Society of Japan **65**, 3056 (1996).

[S9]  L. Caretta *et al.*, Nature Communications **11**, 1090 (2020).

[S10]  G. van der Laan and B. T. Thole, Physical Review B **53**, 14458 (1996).

[S11]  T. Jo, Journal of Electron Spectroscopy and Related Phenomena **86**, 73 (1997).

[S12]  H. B. Vasili *et al.*, Physical Review B **96**, 014433 (2017).

[S13]  A. Goldman, *Modern ferrite technology* (Springer Science & Business Media, 2006).

[S14]  R. K. Wangsness, Physical Review **91**, 1085 (1953).

[S15]  G. P. Rodrigue, H. Meyer, and R. V. Jones, Journal of Applied Physics **31**, S376 (1960).

**References**

[1]  J. H. Van Vleck, *Physical Review* **1950**, 78, 266.

[2]  B. I. Min, Y. R. Jang, *Journal of Physics: Condensed Matter* **1991**, 3, 5131.

[3]  S. Blundell, *Magnetism in condensed matter*, Oxford ; New York : Oxford University Press, 2001., **2001**.

[4]  A. Soumyanarayanan, N. Reyren, A. Fert, C. Panagopoulos, *Nature* **2016**, 539, 509.

[5]  I. Dzyaloshinsky, *Journal of Physics and Chemistry of Solids* **1958**, 4, 241.

[6]  T. Moriya, *Physical Review* **1960**, 120, 91.

[7]  M. Bode, M. Heide, K. von Bergmann, P. Ferriani, S. Heinze, G. Bihlmayer, A. Kubetzka, O. Pietzsch, S. Blügel, R. Wiesendanger, *Nature* **2007**, 447, 190.

[8]  B. A. Bernevig, T. L. Hughes, S.-C. Zhang, *Science* **2006**, 314, 1757.

[9]  C. O. Avci, A. Quindeau, C.-F. Pai, M. Mann, L. Caretta, A. S. Tang, M. C. Onbasli, C. A. Ross, G. S. D. Beach, *Nature Materials* **2017**, 16, 309.

[10]  A. Quindeau, C. O. Avci, W. Liu, C. Sun, M. Mann, A. S. Tang, M. C. Onbasli, D. Bono, P. M. Voyles, Y. Xu, J. Robinson, G. S. D. Beach, C. A. Ross, *Advanced Electronic Materials* **2017**, 3, 1600376.

[11]  C. O. Avci, E. Rosenberg, L. Caretta, F. Büttner, M. Mann, C. Marcus, D. Bono, C. A. Ross, G. S. D. Beach, *Nature Nanotechnology* **2019**, 14, 561.

[12]  S. Vélez, J. Schaab, M. S. Wörnle, M. Müller, E. Gradauskaite, P. Welter, C. Gutgsell, C. Nistor, C. L. Degen, M. Trassin, M. Fiebig, P. Gambardella, *Nature Communications* **2019**, 10, 4750.

[13]  L. Caretta, E. Rosenberg, F. Büttner, T. Fakhrul, P. Gargiani, M. Valvidares, Z. Chen, P. Reddy, D. A. Muller, C. A. Ross, G. S. D. Beach, *Nature Communications* **2020**, 11, 1090.

[14]  L. Caretta, S.-H. Oh, T. Fakhrul, D.-K. Lee, B. H. Lee, S. K. Kim, C. A. Ross, K.-J. Lee, G. S. D. Beach, *Science* **2020**, 370, 1438.

[15]  Q. Shao, Y. Liu, G. Yu, S. K. Kim, X. Che, C. Tang, Q. L. He, Y. Tserkovnyak, J. Shi, K. L. Wang, *Nature Electronics* **2019**, 2, 182.

[16]  Q. Shao, A. Grutter, Y. Liu, G. Yu, C.-Y. Yang, D. A. Gilbert, E. Arenholz, P. Shafer, X. Che, C. Tang, M. Aldosary, A. Navabi, Q. L. He, B. J. Kirby, J. Shi, K. L. Wang, *Physical Review B* **2019**, 99, 104401.

[17]  A. Schreyer, T. Schmitte, R. Siebrecht, P. Bödeker, H. Zabel, S. H. Lee, R. W. Erwin, C. F. Majkrzak, J. Kwo, M. Hong, *Journal of Applied Physics* **2000**, 87, 5443.

[18]  P. Schattschneider, S. Rubino, C. Hébert, J. Rusz, J. Kuneš, P. Novák, E. Carlino, M. Fabrizioli, G. Panaccione, G. Rossi, *Nature* **2006**, 441, 486.

[19]    M. Caminale, A. Ghosh, S. Auffret, U. Ebels, K. Ollefs, F. Wilhelm, A. Rogalev, W. E. Bailey, *Physical Review B* **2016**, 94, 014414.

[20]    S. Tacchi, R. E. Troncoso, M. Ahlberg, G. Gubbiotti, M. Madami, J. Åkerman, P. Landeros, *Physical Review Letters* **2017**, 118, 147201.

[21]    C. N. Wu, C. C. Tseng, K. Y. Lin, C. K. Cheng, S. L. Yeh, Y. T. Fanchiang, M. Hong, J. Kwo, *AIP Advances* **2017**, 8.

[22]    C. N. Wu, C. C. Tseng, Y. T. Fanchiang, C. K. Cheng, K. Y. Lin, S. L. Yeh, S. R. Yang, C. T. Wu, T. Liu, M. Wu, M. Hong, J. Kwo, *Scientific Reports* **2018**, 8, 11087.

[23]    O. Ciubotariu, A. Semisalova, K. Lenz, M. Albrecht, *Scientific Reports* **2019**, 9, 17474.

[24]    J. M. Shaw, H. T. Nembach, T. J. Silva, *Physical Review B* **2013**, 87, 054416.

[25]    D. Cheshire, P. Bencok, D. Gianolio, G. Cibin, V. K. Lazarov, G. v. d. Laan, S. A. Cavill, *Journal of Applied Physics* **2022**, 132, 103902.

[26]    C. Love, J. E. Beevers, B. Achinuq, R. Fan, K. Matsuzaki, T. Susaki, V. K. Lazarov, S. S. Dhesi, G. van der Laan, S. A. Cavill, *Physical Review B* **2023**, 107, 064414.

[27]    H. Elnaggar, P. Sainctavit, A. Juhin, S. Lafuerza, F. Wilhelm, A. Rogalev, M. A. Arrio, C. Brouder, M. van der Linden, Z. Kakol, M. Sikora, M. W. Haverkort, P. Glatzel, F. M. F. de Groot, *Physical Review Letters* **2019**, 123, 207201.

[28]    S. Mokarian Zanjani, M. C. Onbaşlı, *Journal of Magnetism and Magnetic Materials* **2020**, 499, 166108.

[29]    Q. Shao, C. Tang, G. Yu, A. Navabi, H. Wu, C. He, J. Li, P. Upadhyaya, P. Zhang, S. A. Razavi, Q. L. He, Y. Liu, P. Yang, S. K. Kim, C. Zheng, Y. Liu, L. Pan, R. K. Lake, X. Han, Y. Tserkovnyak, J. Shi, K. L. Wang, *Nature Communications* **2018**, 9, 3612.

[30]    R. Zhang, R. F. Willis, *Physical Review Letters* **2001**, 86, 2665.

[31]    F. Huang, M. T. Kief, G. J. Mankey, R. F. Willis, *Physical Review B* **1994**, 49, 3962.

[32]    S. Crossley, A. Quindeau, A. G. Swartz, E. R. Rosenberg, L. Beran, C. O. Avci, Y. Hikita, C. A. Ross, H. Y. Hwang, *Applied Physics Letters* **2019**, 115, 172402.

[33]    C. Hauser, T. Richter, N. Homonnay, C. Eisenschmidt, M. Qaid, H. Deniz, D. Hesse, M. Sawicki, S. G. Ebbinghaus, G. Schmidt, *Scientific Reports* **2016**, 6, 20827.

[34]    C. Kittel, *Physical Review* **1949**, 76, 743.

[35]    J. M. Shaw, H. T. Nembach, T. J. Silva, C. T. Boone, *Journal of Applied Physics* **2013**, 114, 243906.

[36]    C. Gonzalez-Fuentes, R. K. Dumas, C. García, *Journal of Applied Physics* **2018**, 123, 023901.

[37]   M. Farle, *Reports on Progress in Physics* **1998**, 61, 755.

[38]   A. J. P. Meyer, G. Asch, *Journal of Applied Physics* **1961**, 32, S330.

[39]   T. M. Dunn, *Transactions of the Faraday Society* **1961**, 57, 1441.

[40]   D. N. Petrov, B. M. Angelov, *Physica B: Condensed Matter* **2019**, 557, 103.

[41]   M. Blume, S. Geschwind, Y. Yafet, *Physical Review* **1969**, 181, 478.

[42]   K. Gilmore, Y. U. Idzerda, M. D. Stiles, *Journal of Applied Physics* **2008**, 103, 07D303.

[43]   E. Goering, *physica status solidi (b)* **2011**, 248, 2345.

[44]   H. T. Nembach, T. J. Silva, J. M. Shaw, M. L. Schneider, M. J. Carey, S. Maat, J. R. Childress, *Physical Review B* **2011**, 84, 054424.

[45]   A. N. Anisimov, M. Farle, P. Poulopoulos, W. Platow, K. Baberschke, P. Isberg, R. Wäppling, A. M. N. Niklasson, O. Eriksson, *Physical Review Letters* **1999**, 82, 2390.

[46]   A. Barla, J. Nicolas, D. Cocco, S. M. Valvidares, J. Herrero-Martin, P. Gargiani, J. Moldes, C. Ruget, E. Pellegrin, S. Ferrer, *Journal of Synchrotron Radiation* **2016**, 23, 1507.

[47]   J. Lourembam, X. Yu, M. P. R. Sabino, M. Tran, R. W. T. Ang, Q. J. Yap, S. Ter Lim, A. Rusydi, *Physical Review Applied* **2020**, 14, 054022.

[48]   G. Kaindl, G. Kalkowski, W. D. Brewer, B. Perscheid, F. Holtzberg, *Journal of Applied Physics* **1984**, 55, 1910.

[49]   W. Haubenreisser, *Kristall und Technik* **1979**, 14, 1490.

[50]   J. J. van Loef, *Solid State Communications* **1966**, 4, 625.

[51]   J. D. Litster, G. B. Benedek, *Journal of Applied Physics* **1966**, 37, 1320.

[52]   H. B. Vasili, B. Casals, R. Cichelero, F. Macià, J. Geshev, P. Gargiani, M. Valvidares, J. Herrero-Martin, E. Pellegrin, J. Fontcuberta, G. Herranz, *Physical Review B* **2017**, 96, 014433.

[53]   T. S. Herng, W. Xiao, S. M. Poh, F. He, R. Sutarto, X. Zhu, R. Li, X. Yin, C. Diao, Y. Yang, X. Huang, X. Yu, Y. P. Feng, A. Rusydi, J. Ding, *Nano Research* **2015**, 8, 2935.

[54]   F. Frati, M. O. J. Y. Hunault, F. M. F. de Groot, *Chemical Reviews* **2020**, 120, 4056.

[55]   J. Stöhr, H. C. Siegmann, *Magnetism: From Fundamentals to Nanoscale Dynamics, Springer, Berlin*, **2006**.

[56]   C. T. Chen, Y. U. Idzerda, H. J. Lin, N. V. Smith, G. Meigs, E. Chaban, G. H. Ho, E. Pellegrin, F. Sette, *Physical Review Letters* **1995**, 75, 152.

[57]   J. M. Shaw, R. Knut, A. Armstrong, S. Bhandary, Y. Kvashnin, D. Thonig, E. K. Delczeg-Czirjak, O. Karis, T. J. Silva, E. Weschke, H. T. Nembach, O. Eriksson, D. A. Arena, *Physical Review Letters* **2021**, 127, 207201.